\documentclass[twocolumn,prl,showpacs,amsmath,amssymb,superscriptaddress,floatfix,nofootinbib]{revtex4}

\usepackage{graphicx}
\usepackage{dcolumn}
\usepackage{bm}
\usepackage{color}

\begin{document}

\title{
Shape transitions in exotic Si and S isotopes and
tensor-force-driven Jahn-Teller effect
}

\author{Yutaka Utsuno}
\affiliation{Advanced Science Research Center, Japan Atomic Energy
Agency, Tokai, Ibaraki 319-1195, Japan}
\affiliation{Center for Nuclear Study, University of Tokyo, Hongo,
Bunkyo-ku, Tokyo 113-0033, Japan}
\author{Takaharu Otsuka}
\affiliation{Department of Physics, University of Tokyo, Hongo,
Bunkyo-ku, Tokyo 113-0033, Japan}
\affiliation{Center for Nuclear Study, University of Tokyo, Hongo,
Bunkyo-ku, Tokyo 113-0033, Japan}
\affiliation{National Superconducting Cyclotron Laboratory,
Michigan State University, East Lansing MI 48824, USA}
\author{B. Alex Brown}
\affiliation{National Superconducting Cyclotron Laboratory,
Michigan State University, East Lansing MI 48824, USA}
\affiliation{Department of Physics, Michigan State University,
East Lansing MI 48824, USA}
\author{Michio Honma}
\affiliation{Center for Mathematical Sciences, University of Aizu,
Ikki-machi, Aizu-Wakamatsu, Fukushima 965-8580, Japan}
\author{Takahiro Mizusaki}
\affiliation{Institute for Natural Sciences, Senshu University, Tokyo,
  101-8425, Japan}
\author{Noritaka Shimizu}
\affiliation{Center for Nuclear Study, University of Tokyo, Hongo,
Bunkyo-ku, Tokyo 113-0033, Japan}
\date{\today}

\begin{abstract}
We show how shape transitions in the neutron-rich exotic Si and S
isotopes occur in terms of shell-model calculations with a newly 
constructed Hamiltonian based on $V_{\rm MU}$ interaction. 
We first compare the calculated spectroscopic-strength distributions
for the proton $0d_{5/2,3/2}$ and $1s_{1/2}$ orbitals with 
results extracted from a $^{48}$Ca(e,e'p) experiment to show
the importance of the tensor-force component of
the Hamiltonian. Detailed calculations for the excitation energies,
B(E2) and two-neutron separation energies for the Si and S isotopes show excellent
agreement with experimental data. The potential energy surface exhibits
rapid shape transitions along the isotopic chains towards $N$=28
that are different for Si and S.
We explain the results in
terms of an intuitive picture involving a Jahn-Teller-type
effect that is sensitive to the tensor-force-driven shell evolution.
The closed sub-shell nucleus $^{42}$Si is a particularly good
example of how the tensor-force-driven Jahn-Teller mechanism
leads to a strong oblate rather than spherical shape.
\end{abstract}


\maketitle
Among the new frontiers of nuclear physics, one of the most
important is the evolution of single-particle energies
in nuclei far from stability that led to significant variation of
the shell structure and even dramatic changes in the
location of magic numbers
\cite{review,magic}.
One of the ingredients is the tensor force which can change spin-orbit
splitting significantly
leading to tensor-force driven shell evolution \cite{tensor,vmu}.
While many experimental examples, {\it e.g.}, 
\cite{schiffer,gaudefroy,75Cu,101Sn},
have been accumulated on this phenomenon,
its direct test including fragmentation of
single-particle strength has not been presented. 
We shall show, in this paper, the first test of this kind for
proton $0d_{5/2}$-$0d_{3/2}$ 
splitting in $^{48}$Ca, comparing to spectroscopic factors
measured in the (e,e'p) experiment \cite{kramer}.

We shall then show\footnote{A brief account 
with selected results of this work was presented in \cite{inpc}.} 
that tensor-driven shell evolution plays
a critical role in the rapid shape transition
as a function of neutron and/or proton number, including triaxial and
$\gamma$-unstable shapes. 
In particular, we show for the first time 
how this shape transition 
at low energy can be related to the Jahn-Teller type
effect \cite{jt, reinhard}, where
a geometric distortion is brought about by a particular coherent
superposition of relevant single-particle states enhanced due to
their (near) degeneracy.

These studies are performed in terms of the shell model in a
unified way with a new Hamiltonian.
We shall show calculated energy levels and B(E2) values
for the Si and S isotopes that are in 
good agreement with experiments 
\cite{ensdf, si36, si40, si42, s40, s42, s44},
and new predictions are made.
Potential energy surfaces (PES) 
are shown.
The PES changes rapidly as a function of neutron number, and are
different for Si and S.
The PES for $^{42}$Si shows a strong oblate shape
(rather than spherical)
due to the tensor-force-driven shell 
evolution.
These PES results are interpreted in an intuitive picture involving
shell gaps and the Jahn-Teller-type effect.
Two-neutron separation energies are discussed also.

We outline the present shell-model calculations.
The $sd$ and $pf$ shells are taken as the valence shell with protons
in $sd$ and neutrons in $pf$.
The interactions within each of these shells are based on existing
interactions: USD \cite{usd} 
(GXPF1B \cite{gxpf1b}\footnote{GXPF1B Hamiltonian was created 
from GXPF1A Hamiltonian \cite{gxpf1a} by changing five 
$T=1$ matrix elements and SPE involving $1p_{1/2}$.  Such 
differences give no notable change to the present work due to 
minor relevance of $1p_{1/2}$.})
for the $sd$ ($pf$) shell, except for
the monopole interactions \cite{vmu,kb3} 
$V^{T=0,1}_{0d_{3/2},0d_{5/2}}$ based on SDPF-M \cite{sdpf-m}
due to a problem in USD as pointed out in \cite{magic}.  
The monopole- and quadrupole-pairing matrix elements
$\langle 0f_{7/2} 0f_{7/2} \left| V \right| 0f_{7/2} 0f_{7/2}\rangle_{J=0,2}$
are replaced with those of KB3 \cite{kb3}.  This is mainly
for a better description of nuclei of $N\sim$ 22.     
Although this replacement is not so relevant to the present study
where $N$ is larger in most cases, it has been made so that
the applicability of the new Hamiltonian becomes wider.
The cross-shell part, most essential for exotic nuclei discussed
in this study but rather undetermined so far,
is given basically by $V_{\rm MU}$ of \cite{vmu} with small
refinements stated below.
The tensor-force component is exactly from $V_{\rm MU}$, implying
the $\pi + \rho$ meson exchange force.  
It has been used in many cases \cite{tensor, vmu}, and
it has been accounted for microscopically under the new concept of
Renormalization Persistency \cite{tsunoda} where modern
realistic effective interactions are analyzed in terms of the
spin-tensor decomposition technique \cite{spintensor1,spintensor2,smir}.
The central-force component 
of $V_{\rm MU}$ interaction has been determined 
in \cite{vmu} so as to reproduce monopole properties of 
shell-model interactions SDPF-M ($sd$-shell part)
and GXPF1A \cite{gxpf1,gxpf1a} by a simple Gaussian interaction.  
We introduce here slight fine tuning with density dependence 
similar to the one in \cite{spintensor2},
in order that its monopole part becomes closer to that of
GXPF1B.
We include the two-body spin-orbit force of the M3Y
interaction \cite{m3y}.
(For the present study, all these refinements produce minor
changes, and do not change the overall conclusions.
The refinements
have been made so that the new Hamiltonian works well
in a wide variety of nuclei other than those of this study.)
Following USD and GXPF1B,
all two-body matrix elements are scaled by $A^{-0.3}$.
The single-particle energies (SPE) of $sd$ shell are taken
from USD, and those of $pf$ shell are determined by requesting
their effective SPEs  
on top of $^{40}$Ca closed shell equal to the single-particle
energies of GXPF1B.

The Hamiltonian, referred to as 
SDPF-MU hereafter, 
is thus fixed prior to the shell model calculations
presented in this paper.
The diagonalization is performed 
by the {\sc mshell64} code \cite{mshell}.
The $V_{\rm MU}$ interaction has been used also to construct
the cross-shell part of a recent shell-model Hamiltonian
for $p$-$sd$ shell nuclei including neutron-rich exotic ones,
providing with a good description of very light (B,C,N,O) nuclei 
\cite{yuan}.  

\begin{figure}[tb]
 \begin{center}
 \includegraphics[width=8.5cm,clip]{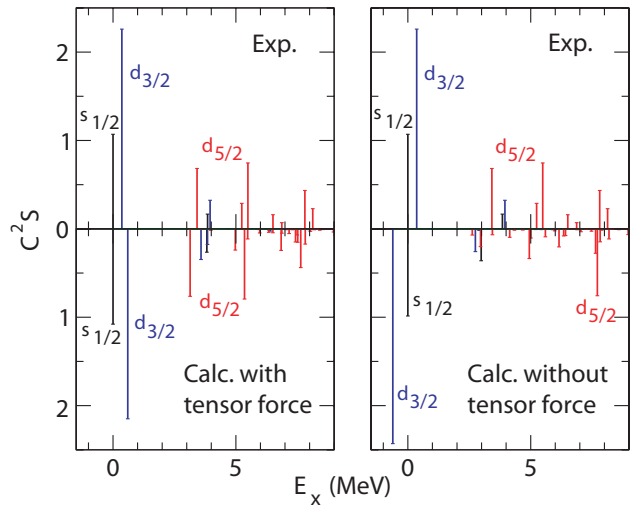}
 \caption{
(Color online) Spectroscopic factors of proton hole states
measured by $^{48}$Ca(e,e'p) \cite{kramer} (upper) and
its theoretical calculation (lower left).  The cross-shell tensor force
is removed in lower right panel.    
The black, blue and red bars correspond to $1s_{1/2}$, $0d_{3/2}$ and
$0d_{5/2}$ states, respectively.
}
 \label{fig:sfac}
 \end{center}
\end{figure}

We begin with the distribution of single-particle strength
of proton $sd$-shell orbits.
Spectroscopic factors obtained for $^{48}$Ca
with the ($e, e'p$) reaction are displayed in the upper panels
of Fig.~\ref{fig:sfac} \cite{kramer}.
The $0d_{5/2}$ single-particle strength is highly fragmented
due to its high excitation energy (3-8 MeV range).
The spectroscopic factors obtained by the present calculation
are shown in the lower left-hand panel of Fig.~\ref{fig:sfac},
where an overall quenching factor 0.7 is used following the
standard recipe to incorporate various effects of components
outside the valence shell \cite{barbieri}.
The agreement is excellent both in the position of peaks and their
magnitudes. 
However, this agreement is lost, if the tensor force is removed
from the cross-shell interaction, as shown in the right lower
panel of Fig.~\ref{fig:sfac}.  For instance, the largest $0d_{3/2}$
and  $1s_{1/2}$ peaks are in the wrong order, and the strongest peaks of
$0d_{5/2}$ move towards higher energy.
The $0d_{3/2}$-$0d_{5/2}$ gap of $^{48}$Ca turns out to be 
$\sim$5 MeV in the present calculation, but becomes $\sim$2 MeV
larger, 
if the cross-shell tensor force is switched off.

For $^{40}$Ca, although no experimental data by ($e, e'p$) reaction is
available \cite{kramer89}, Bastin {\it et al.} suggested
a reduction of proton $0d_{5/2}$-$0d_{3/2}$ gap by 1.9 MeV
from $^{40}$Ca to $^{48}$Ca based on reaction data \cite{si42}.
Effective SPEs obtained from the SDPF-MU Hamiltonian are consistent
with this and other experimental data \cite{review}.
On the other hand, the gap remains almost unchanged between $^{40}$Ca
and $^{48}$Ca, if the cross-shell tensor force is removed. 
We note that the spectroscopic factor distributions for $0d_{3/2}$
and $1s_{1/2}$ were calculated also by a Green's function method
by Barbieri \cite{barbieri}, 
but there has been no previous report on the $0d_{5/2}$ strength.

The proton shell structure thus evolves from $^{40}$Ca to $^{48}$Ca.
Because only the tensor force can change the $0d_{3/2}$-$0d_{5/2}$
gap by this order of magnitude ($\sim$2 MeV),
the agreement shown in Fig.~\ref{fig:sfac}
provides us with the first evidence from electron scattering
experiments to the tensor-force-driven shell evolution
induced by the mechanism of Otsuka {\it et al.} \cite{tensor}.
This agreement implies also the validity of the present SDPF-MU
Hamiltonian, especially the interaction between the proton $sd$
and neutron $pf$ shells.

\begin{figure}[tb]
 \begin{center}
 \includegraphics[width=7cm,clip]{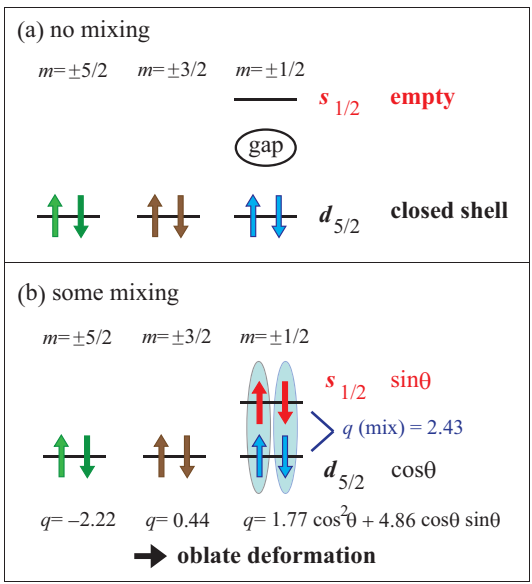}
 \caption{
(Color online) Intuitive illustration of the structure of intrinsic
state at $Z$=14.  Single-particle states of magnetic quantum numbers,
denoted $m$, are shown.  $q$ implies intrinsic quadrupole moment
(fm$^2$) obtained with harmonic oscillator $\hbar \omega$=11.8 MeV.
}
 \label{fig:oblate}
 \end{center}
\end{figure}

We now move to the shape transitions in exotic Si ($Z$=14) and
S ($Z$=16) isotopes with even $N$=22-28.
Before discussing quantitative results,
we present an intuitive picture in
Fig.~\ref{fig:oblate} on the relation between the shell
structure and the shape of Si nuclei.   
To begin with,
we consider only two orbits $0d_{5/2}$ and $1s_{1/2}$ of protons
for the sake of simplicity. 
In a conventional view, $Z$=14 is a sub-magic
number : no mixing between $1s_{1/2}$ and $0d_{5/2}$ due to
large $1s_{1/2}$-$0d_{5/2}$ gap and/or weak mixing force.
Six protons occupy all states of $0d_{5/2}$ forming a
closed subshell, as depicted in Fig.~\ref{fig:oblate}(a).
This should end up with a spherical shape for a doubly magic
nucleus, $^{42}$Si ($N$=28), similarly to $^{34}$Si ($N$=20).

\begin{figure}[tb]
 \begin{center}
 \includegraphics[width=8.5cm,clip]{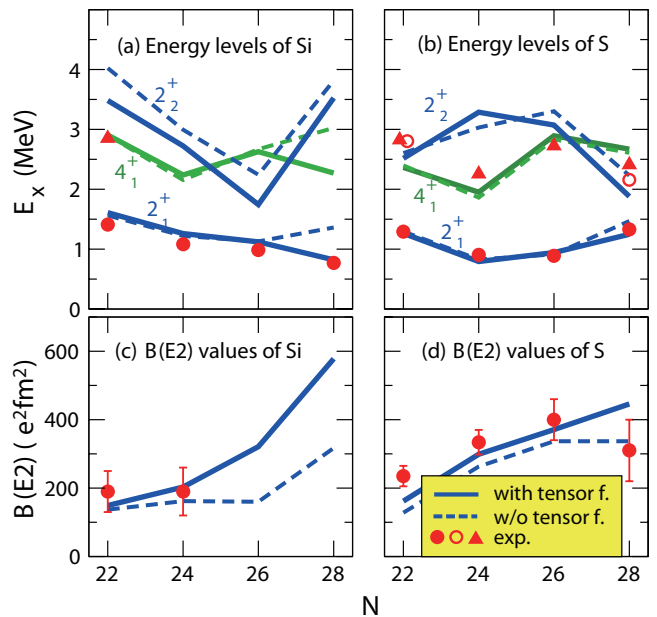}
 \caption{
(Color online)
(a,b) 2$^+_{1,2}$ (blue lines and red circles) and 4$^+_1$ (green lines
  and red triangles) energy levels and (c,d)
 $B$(E2;0$^+_1$$\to$2$^+_1$) values of Si and S isotopes
for $N$=22-28.
Symbols are experimental data
\cite{ensdf, si36, si40, si42, s40, s42, s44}.
Solid (dashed) lines are calculations
with (without) the cross-shell tensor force.
}
 \label{fig:si-s}
 \end{center}
\end{figure}

Figure~\ref{fig:oblate} (b) indicates another situation where
sizable mixing occurs between the $1s_{1/2}$ and $0d_{5/2}$ orbits
in the case when the proton-neutron correlation is stronger than
in Fig.~\ref{fig:oblate} (a) and competes with 
$1s_{1/2}$-$0d_{5/2}$ spacing. 
The proton-neutron interaction, apart from its monopole part, can be
modeled by a quadrupole-quadrupole interaction to a good extent.
Effects of this interaction can be discussed in terms of intrinsic
states due to a quadrupole deformation.
Assuming an axially symmetric deformation, single-particle states
of the same magnetic quantum numbers, denoted $m$, are
mixed in the intrinsic states.  
Figure~\ref{fig:oblate} (b) shows that this occurs for $m=\pm 1/2$
between $1s_{1/2}$ and $0d_{5/2}$,
with amplitudes sin$\theta$ and cos$\theta$, respectively. 
The phase of the mixing amplitude depends on the shape,
prolate or oblate.
In the case of Si isotopes, protons occupy the
states of $m=\pm 5/2, \pm 3/2$, which yield in total a negative
intrinsic quadrupole moment (oblate). 
The total intrinsic quadrupole moment gains a larger magnitude,
if the $1s_{1/2}$-$0d_{5/2}$ mixing gives a negative moment.
The contribution from the proton-neutron multipole interaction
to the energy of total intrinsic state is proportional
approximately to the
product of the proton intrinsic quadrupole moment and the neutron one.
Because a similar situation can occur for neutrons in $1p_{3/2}$ and
$0f_{7/2}$ with a negative intrinsic moment produced mainly by
the $m=\pm 7/2$ component of $0f_{7/2}$,   
a stronger binding is obtained for the total intrinsic state
with an oblate shape.
Although the proton $0d_{3/2}$ orbit is mixed to some extent 
in the actual shell-model calculation, the above mechanism
still holds :
the occupation of $m=\pm 5/2$ remains with large negative
quadrupole moment (oblate), 
and the mixing can occur among $0d_{5/2}$, $0d_{3/2}$ and
$1s_{1/2}$ orbits in the relevant $m=\pm 1/2$ and $\pm 3/2$
components in favor of the oblate shape.

This mechanism is of Jahn-Teller type, and it works 
if the mixing due to the proton-neutron correlation
can compete with SPE spacings. 
The above intuitive pictures with Figs.~\ref{fig:oblate} (a,b)
suggest rather robustly that the shape of
$^{42}$Si can be spherical or oblate, but not prolate.
The $0d_{5/2}$-$1s_{1/2}$ spacing is 7.8 MeV for $^{42}$Si
with the SDPF-MU Hamiltonian in the filling scheme.  This is
indeed comparable with the ground-state expectation value,
-13.2 MeV, of multipole proton-neutron interaction,
obtained by the shell-model calculation discussed below.  
Regarding the tensor force,
when many neutrons occupy the $0f_{7/2}$ orbital, the proton
$0d_{5/2}$ is raised
due to the mechanism of Otsuka {\it et al.} \cite{tensor}
discussed for the (e,e'p) data above. 
This effect is included in the result of SDPF-MU Hamiltonian,
reducing $0d_{5/2}$-$1s_{1/2}$ spacing by 1.1 MeV for
$^{42}$Si. We shall see its importance now.

\begin{figure}[t]
 \begin{center}
 \includegraphics[width=7.5cm,clip]{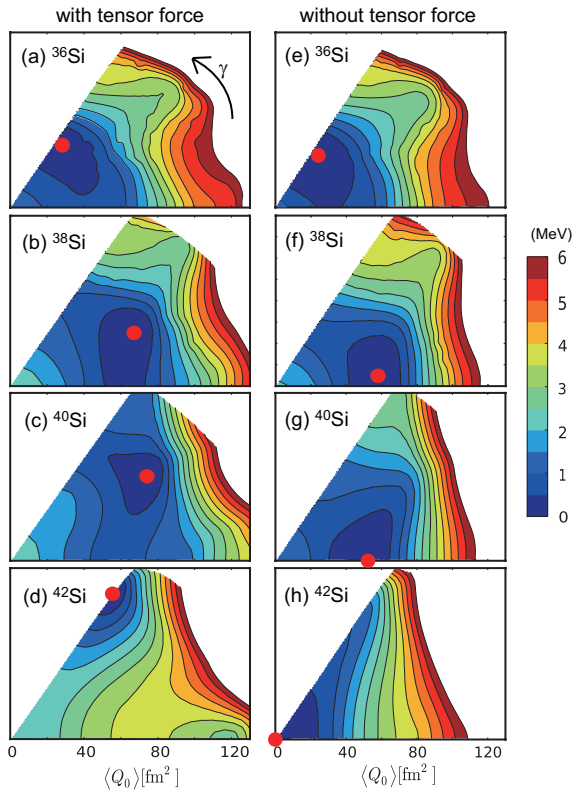}
 \caption{
(Color online)
Potential energy surfaces of Si isotopes for 
$\gamma$=0$\sim$60$^{\circ}$ from $N=22$ to 28
calculated with (left) and without (right) the cross-shell
tensor force. The energy minima are indicated by red circles.
}
 \label{fig:pes}
 \end{center}
\end{figure}

\begin{figure}[t]
 \begin{center}
 \includegraphics[width=7.5cm,clip]{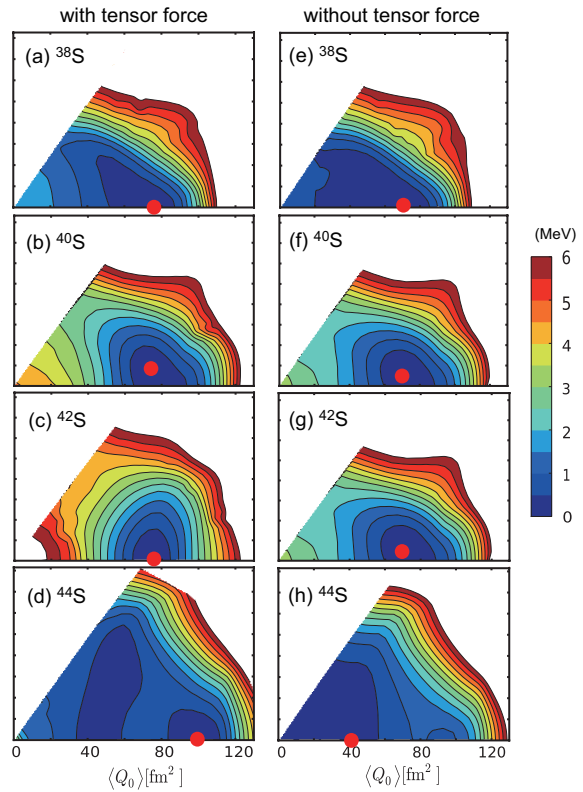}
 \caption{
(Color online)
Potential energy surfaces of S isotopes from $N=22$ to 28.
See the caption of Fig. \ref{fig:pes}.
}
 \label{fig:pes_S}
 \end{center}
\end{figure}

We here investigate quantitatively the structure of Si and S
isotopes in the context of the shell-model calculations with the
SDPF-MU Hamiltonian.  Figure \ref{fig:si-s} exhibits 
properties of even-$A$ Si and S isotopes.
Effective charges are $(e_p, e_n)=(1.35e, 0.35e)$ where
an isoscalar shift fixed for lighter isotopes are taken.
The overall agreement with experiment is excellent in
Fig. \ref{fig:si-s}.
For instance, in the present result,
$2^+_1$ levels of Si isotopes keep coming down as $N$ increases
consistently with experiment.
The nice agreement suggests that the intuitive picture with
Fig.~\ref{fig:oblate} (b)
works particularly well towards $^{42}$Si, resulting in a
strongly deformed shape with low excitation energies
consistent with recent measurement in GANIL \cite{si42}. 
However, if the tensor force is omitted from the cross-shell
interaction, the $2^+_1$ level of $^{42}$Si goes up, suggesting
the case in Fig.~\ref{fig:oblate} (a).
Figure \ref{fig:si-s} exhibits results for S isotopes also
in good overall agreement, including a bumpy behavior of the
$4^+_1$ level.
Earlier shell-model calculations 
with empirical interactions
\cite{sdpf-u,kaneko}
give larger deviations and/or different trends from experiments.
Different Hamiltonians are taken in \cite{sdpf-u} 
between $Z \le$14 and $Z \ge$15 isotopes related to
the monopole pairing strength in $pf$ shell.
The deviation from experiment becomes larger if this
change is switched off.
The present Hamiltonian is the same for all isotopes, and has been
fixed prior to the shell-model calculations so as to make
predictions.  

The potential energy surface (PES) can be used to understand shapes
contained in theoretical calculations.  Figures~\ref{fig:pes} and
\ref{fig:pes_S} exhibit PES for Si and S isotopes, respectively,  
obtained by the constrained Hartree-Fock method \cite{ni56} for the
SDPF-MU Hamiltonian.  The full Hamiltonian is taken in panels
(a$\sim$d) of the two figures, whereas the cross-shell tensor force
is removed in panels (e$\sim$h).
We begin with PESs of Si isotopes (Fig.~\ref{fig:pes}). 
Shape evolutions are seen clearly in both sequences
(a$\sim$d) and (e$\sim$h), starting with similar patterns in $^{36}$Si.
The shape evolves as more neutrons occupy $pf$-shell, with distinct
differences between the two sequences.
In (b,c), the deformation becomes stronger from (a) with
triaxial minima, whereas the shape becomes more like modestly prolate
in (f,g).  In (d), one finds a strongly oblate shape with
a sharp minimum, but the minimum is at the spherical shape in (h).
This strong oblate deformation produces low 2$^+$ level and
large $B$(E2) in Fig.~\ref{fig:si-s} for the ``doubly-closed''
$^{42}$Si.
Thus, the shape of exotic Si isotopes changes significantly within
the range of $\Delta N \sim$6.  This feature is partly due to
growing collectivity with more neutrons in the $pf$ shell, but
is also a manifestation of Jahn-Teller-type effect driven by the
tensor force.  Without the tensor force, the SPE spacings are
too large, and the correlation energies cannot produce this effect.

The $\gamma$-unstable deformation is well developed
in Fig.~\ref{fig:pes} (c), and this can be confirmed by the
low-lying $2^+_2$ level of $^{40}$Si in Fig.~\ref{fig:si-s}.
This level seems to agree with a recent $\gamma$-ray experiment
\cite{si40} where either of $\gamma$-rays
638(8) and 845(6) keV appears to feed directly the $2^+_1$ state.
We stress that the $2^+_2$ level is sensitive to the
tensor force through $\gamma$-instability in Si isotopes.
In fact, the ratio $E_x(2^+_2) / E_x(2^+_1)$ is 
as low as 1.5 for $^{40}$Si, whereas it becomes 4.4 for $^{42}$Si.
The former is a prominent signature of $\gamma$-instability, 
while the latter is consistent with a vibration from 
a profound PES minimum of axially-symmetric deformation.
Thus, the change from $^{40}$Si to $^{42}$Si is an intriguing
example of the rapid and unexpected structure evolution.
If the tensor force is switched off in the cross-shell
interaction, $E_x(2^+_1)$ of $^{42}$Si is raised but $B$(E2) value
is still larger than that of $^{40}$Si.  This is partly due to
the stretched minimum in Fig.~\ref{fig:pes} (h) and partly 
due to relatively enhanced proton contribution in the E2 
transition because of $N$=28 closure.

The situation is quite different with PES of S isotopes
(Fig.~\ref{fig:pes_S}).
With two more protons added to Si isotopes, the occupancy of the $0d_{5/2}$ orbit
is closer to that of a closed shell.
The mixing between $1s_{1/2}$ and $0d_{3/2}$
then plays a decisive role, leading to a structure favoring prolate
deformation.
As for neutrons, $m=\pm 7/2$ components of $0f_{7/2}$ carry
large negative intrinsic
quadrupole moment and are crucial for oblate shape,
while the others have positive or small negative values.    
In S isotopes, as protons favor prolate shapes,
neutron sector becomes prolate for $N<$28, by keeping
the $m=\pm 7/2$ states (almost) unoccupied.
At $N$=28, the $m=\pm 7/2$ states are occupied more.  However,
the neutron $0f_{7/2}$-$1p_{3/2}$ spacing decreases from Ca isotopes
to Si and S ones due to the vacancy of proton $0d_{3/2}$, and
the excitation from the $m=\pm 7/2$ states to
other orbits does not cost much energy as compared to gains from 
proton-neutron correlation.  This gives rise not to an oblate
minimum but to triaxial softness.
Thus, the shape is determined, in the present cases,
primarily by the proton sector.
We would like to emphasize the crucial role of 
protons in determining shapes of Si and S isotopes.  
We comment that the quasi SU(3) scheme \cite{quasi},
constructed for the configurations comprised of 
almost degenerate $j$ and $j-2$ orbits only, 
{\it e.g.}, $0f_{7/2}$ and $1p_{3/2}$, 
favors an oblate shape at $N$=28.

Panels (d) and (h) in Fig.~\ref{fig:pes_S} show prolate minima with
opposite trends from panels (c,g) which are rather similar to each
other.  This difference arises within prolate
shapes with shallow minima, and thereby its appearance
in the energy levels is modest
(See Fig.~\ref{fig:si-s}).  The
difference of the shapes between Si and S isotopes  
may suppress transfer reactions,
for instance, two-proton removal from $^{44}$S \cite{msu_S}.

\begin{figure}[t]
 \begin{center}
 \includegraphics[width=6.0cm,clip]{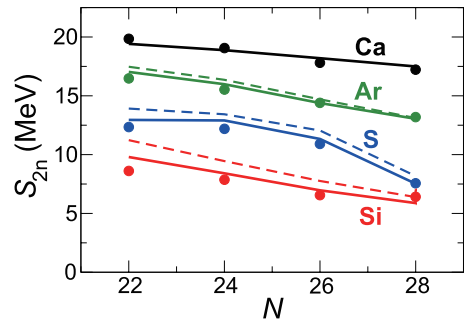}
 \caption{
(Color online)
Two neutron separation energies of Si and S isotopes from $N=22$ to 28.
Solid (dashed) lines are calculations
with (without) the cross-shell tensor force.
Points are experimental data \cite{ame2003, Jurado}.
}
 \label{fig:S2n}
 \end{center}
\end{figure}

We now discuss two-neutron separation energies ($S_{\rm 2n}$) shown
in Fig.~\ref{fig:S2n}.
The agreement between experiment and the full calculation is quite
good within $\sim$0.5 MeV, except for $N$=22 Si value with discrepancy
of 1.2 MeV due to the mixing of intruder configurations
in $^{34}$Si, an issue outside this work.
If the tensor force is switched off, deviations of 1-3 MeV
occur for Si and S isotopes in the direction of larger $S_{2n}$ values. 
This means that the tensor-force effect is repulsive and becomes larger 
from Ca to Si isotopes as a whole, consistently with
the tensor-force-driven shell evolution.
This evolution induces stronger deformation in some cases,
{\it e.g.}, $^{42}$Si, where additional binding energy is gained and can
cancel partly this repulsive effect.
 
While Si and S isotopes have been discussed with density-functional
methods including appearance of oblate shapes
\cite{Lalazissis,Peru,Rodriguez,Li}, there are problems to be
solved.  No systematic calculations have been reported for levels
and B(E2)'s of Si isotopes.
The 2$^+_1$ level calculated in \cite{Li} reproduces experiment 
for $^{44}$S, but deviates by a factor of two for $^{42}$Si.
Fig.~\ref{fig:si-s} (a,b) shows that the 2$^+_1$ level of  
$^{42}$Si is sensitive to the tensor force, whereas that of
$^{44}$S is not.  The difference between $^{42}$Si and $^{44}$S
in ~\cite{Li} might be relevant to this.
The systematics of 2$^+_1$ level in S isotopes shows opposite
trend in \cite{Rodriguez}.   
It will be also of interest to see single-particle properties
given by these works.
For instance, the proton $0d_{5/2}$-$0d_{3/2}$ gap remains 
almost unchanged from $^{40}$Ca to $^{48}$Ca in these calculations,
but this contradicts the trend seen in (e,e'p) data \cite{kramer}.

In summary, we have discussed
the tensor-force driven reduction of the spin-orbit
splitting by the mechanism of Otsuka {\it et al.} \cite{tensor},
for the first time, 
in terms of distribution of spectroscopic factors measured by
$^{48}$Ca(e,e'p) experiment \cite{kramer}.
The SDPF-MU Hamiltonian has been introduced based on the $V_{\rm MU}$
interaction.
The spectroscopic strength distribution
provides a stringent test of this Hamiltonian.
The levels, B(E2)'s and $S_{\rm 2n}$ of exotic Si and S isotopes are described
by the same Hamiltonian in a good agreement with all known experiments,
exhibiting a rather rapid change, as a function
of $N$, for Si isotopes but a quite different change for S isotopes.
The tensor-force driven shell evolution \cite{tensor} 
plays a
crucial role in those shape transitions through the Jahn-Teller-type
effect, including a robust mechanism that favors stable oblate
shapes at sub-shell closures like $^{42}$Si. 
The $B$(E2) values are sensitive to the tensor force. 
The next region of the nuclear chart where such oblate shapes may occur
is near $^{78}$Ni.

This work was in part supported by MEXT Grant-in-Aid for
Scientific Research~(A) 20244022 and for Young Scientists (B) (21740204)
and NSF grant PHY-1068217.
This work
is a part of the CNS-RIKEN joint research project on large-scale
nuclear-structure calculations.


\end{document}